\documentclass[%
 reprint,
superscriptaddress,
 amsmath,amssymb,
 aps,
]{revtex4-2}

\usepackage{graphicx}
\usepackage{dcolumn}
\usepackage{bm}

\usepackage{xcolor}

\begin{document}

\preprint{APS/123-QED}

\title{Laser Excitation of the 1\textit{S}~--~2\textit{S} Transition in Singly-Ionized Helium}

\author{E. L. Gr\"undeman}
\author{V. Barb\'e}%
\author{A. Mart\'inez de Velasco}
\author{C. Roth}
\author{M. Collombon}
\author{J. J. Krauth}
\author{L. S. Dreissen}
\affiliation{LaserLaB, Department of Physics and Astronomy, Vrije Universiteit, De Boelelaan 1081, 1081 HV Amsterdam, the Netherlands}
\author{R. Ta\"ieb}
\affiliation{Sorbonne Universit\'e, CNRS, Laboratoire de Chimie Physique-Mati\'ere et Rayonnement, LCPMR, F-75005 Paris Cedex 05, France
}
\author{K. S. E. Eikema}
\affiliation{LaserLaB, Department of Physics and Astronomy, Vrije Universiteit, De Boelelaan 1081, 1081 HV Amsterdam, the Netherlands}



\date{\today}

\begin{abstract}

Precision laser spectroscopy in the extreme ultraviolet of the 1\textit{S}~--~2\textit{S} two-photon transition in singly-ionized helium is a promising route for tests of fundamental physics. We demonstrate laser excitation of this transition in an atomic beam of $^3$He, based on an amplified frequency comb pulse at 790\,nm combined with its 25$^{\text{th}}$ harmonic at 32\,nm. A clear resonance is observed with a maximum excitation probability of close to 10$^{-4}$ per pulse, and the results are well described by our simulations. This paves the way for high-precision Ramsey-comb spectroscopy of a single helium ion in a Paul trap.

\end{abstract}

\maketitle



Precision spectroscopy of simple, calculable atomic and molecular systems is instrumental for tests of bound-state quantum electrodynamics (QED) and searches for physics beyond the standard model~\cite{Lamb1947,Bethe1947,biraben2009,parthey2011improved,pohl2010size,PatraHD,Ahmadi2018}. 
In order to compare experimental results with theoretical calculations, the fundamental constants are needed as an input to the theory. By combining different measurements, the determination of these fundamental constants can be disentangled from testing the laws of physics. This has sometimes led to surprising results. Up until 2010, the proton charge radius and Rydberg constant were largely based on spectroscopy of hydrogen~\cite{CODATA2006}. However, spectroscopy of muonic hydrogen~\cite{pohl2010size,Pohl2016deuterium}, where the electron is replaced by a muon, resulted in a 4\% adjustment of the proton charge radius and a re-evaluation of the Rydberg constant, as documented in CODATA 2018~\cite{CODATA2018}. Some of the recent measurements performed in normal (electronic) hydrogen agree~\cite{1S3SMPQ,Hesselslambshift,beyer2017} with the updated CODATA value of the proton radius, while others do not \cite{1S3SParis,2S8D}. In order to obtain a consistent picture of the laws of physics and the fundamental constants, it is therefore important to investigate multiple systems.\\
\indent High-precision spectroscopy of the 1\textit{S}~--~2\textit{S} transition in hydrogenlike helium (He$^+$) will enable a new route for tests of fundamental physics~\cite{krauth2019paving,herrmann2009feasibility,moreno2023}. He$^+$ is more sensitive than hydrogen to higher-order QED effects, as these scale with high powers of the nuclear charge. Furthermore, He$^+$ can be confined in an ion trap for long interrogation times and it can be sympathetically cooled by laser-cooled Be$^+$ ions to reduce Doppler effects~\cite{roth2005}. Combining spectroscopy of the 1\textit{S}~--~2\textit{S} transition in He$^+$ with the recently improved determination of the alpha and helion particle charge radii in muonic helium ions~\cite{krauth2021muonichelium,thecremacollaboration2023helion}, and with the Rydberg constant $R_\infty$ from hydrogen spectroscopy will enable stringent tests of higher-order QED terms~\cite{karshenboim2019lamb,yerokhin2019theory}. Alternatively, it could be used to improve the determination of the alpha particle charge radius or to perform a measurement of the Rydberg constant that is nearly independent of hydrogen~\cite{krauth2021muonichelium}.\\
\indent The higher nuclear charge of He$^+$ shifts the 1\textit{S}~--~2\textit{S} two-photon transition wavelength from 2$\times$243\,nm in hydrogen to 2$\times$61\,nm in He$^+$. This is in the extreme ultraviolet (XUV) spectral range where no continuous laser sources exist. Generating enough coherent XUV light to optically excite and perform spectroscopy on this transition is a significant experimental challenge. A promising route towards that goal is the upconversion of frequency comb (FC) laser pulses~\cite{frequency_comb, Jones2000} through high-harmonic generation (HHG)~\cite{mcpherson_studies_1987,corkum_plasma_1993,lewenstein_theory_1994,zerne_phase-locked_1997,bellini_temporal_1998,bartels_generation_2002,guo2018}. Coherent XUV production through HHG has been demonstrated at the full repetition rate of FC lasers in a resonator~\cite{XUVcomb2005,XUVcombYe,nauta2021xuv,moreno2023} for direct FC spectroscopy, and with two amplified FC pulses for Ramsey-comb spectroscopy (RCS)~\cite{morgenweg2014ramsey,RobertH2}. We previously demonstrated RCS combined with HHG and obtained a relative accuracy of 2.3$\times$10$^{-10}$ on the frequency determination of a transition at 110\,nm in xenon~\cite{RCSLaura}, which is the highest accuracy achieved to date with HHG-upconverted light.\\
\indent In this letter, we report on laser excitation of the 1\textit{S}~--~2\textit{S} transition in He$^+$. We produce the He$^+$ ions by photoionization of helium atoms in an atomic beam and excite the 1\textit{S}~--~2\textit{S} transition with a combination of 790\,nm and 32\,nm photons derived from an amplified, HHG-upconverted FC laser pulse. This unequal-wavelength scheme enhances the excitation probability by several orders of magnitude compared to a 2$\times$61\,nm scheme and is compatible with precision RCS. In the current experiment, we use a single laser pulse to study the 1\textit{S}~--~2\textit{S} excitation process in an atomic beam. In this situation the transit time of the helium ions through the XUV interaction zone is too short to interact with the two laser pulses required for RCS. Therefore, future precision RCS spectroscopy will be performed on He$^+$ ions confined in a Paul trap.\\
\begin{figure}[b]
\includegraphics[width=0.45\textwidth]{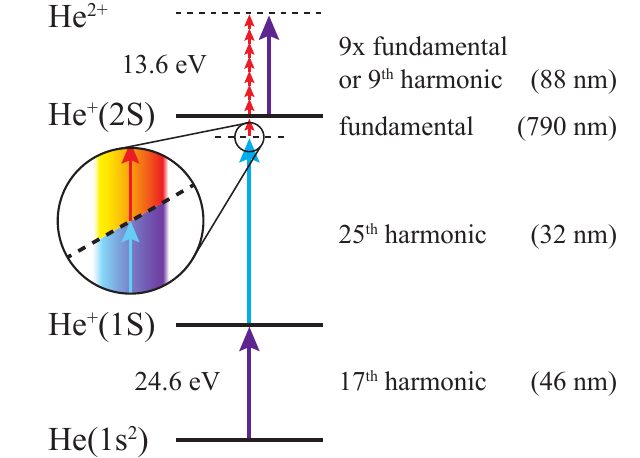}
\caption{\label{fig:helium_levels} Schematic of the multi-photon He$^+$ 1\textit{S}~--~2\textit{S} excitation and detection scheme. One 46\,nm photon (17$^\text{th}$ harmonic of 790\,nm) first ionizes ground-state helium. The 1\textit{S}~--~2\textit{S} in He$^+$ is then driven with one 32\,nm and one 790\,nm photon, and nine 790\,nm photons are used to ionize He$^+$ in the 2S state to He$^{2+}$ (with a small contribution from the 9$^{\text{th}}$ harmonic at 88\,nm). Singly- and doubly-ionized helium are both detected. Zoom-in: many different wavelength combinations add up to the 1\textit{S}~--~2\textit{S} energy difference.}
\end{figure}
%
%
\indent The principle of our He$^+$ generation and excitation scheme is depicted in Fig.~\ref{fig:helium_levels}. All wavelengths are derived from a near-infrared (NIR) amplified FC laser pulse near 790\,nm and its harmonics produced through HHG. Within a single laser pulse, neutral atoms are ionized to He$^+$ by photons of the 17$^\text{th}$ harmonic at 46\,nm, and then resonantly driven from the 1\textit{S} to the 2\textit{S} state with one photon from the fundamental beam at 790\,nm and one from the 25$^{\text{th}}$ harmonic at 32\,nm. The excitation is then detected by state-selective double ionization: only He$^+$ that has been excited to the 2\textit{S} state is further ionized to He$^{2+}$, predominantly by nine-photon ionization at 790\,nm. The entire process from He atoms to He$^{2+}$ ions takes place within the time that higher harmonics are generated, roughly the central {30\,fs of the 150\,fs long NIR pulse. The  He$^{2+}$ ion signal relative to He$^{+}$ signal is a measure of the 1\textit{S}~--~2\textit{S} excitation fraction.\\
\indent The unequal wavelength excitation scheme of 790\,nm and 32\,nm enhances the 1\textit{S}~--~2\textit{S} transition rate by nearly two orders of magnitude compared to 2$\times$61\,nm, because the 32\,nm photon is closer in frequency to the 1\textit{S}~--~2\textit{P} dipole-allowed transition. Moreover, the NIR beam is typically 4--6 orders of magnitude more intense than the generated harmonics, boosting the transition rate by a similar factor. The NIR and 25$^{\text{th}}$ harmonic pulses have a broad bandwidth of about 6 and 20\,THz respectively, compared to the 1\textit{S}~--~2\textit{S} transition linewidth of 84\,Hz. As shown in Fig.~\ref{fig:helium_levels}, for a two-photon transition many different wavelength combinations within the bandwidths of the pulses can equal the 1\textit{S}~--~2\textit{S} energy difference, and all contributions add up coherently to enhance the excitation probability~\cite{baklanov1977}.\\
\begin{figure*}
\includegraphics[width=0.95\textwidth]{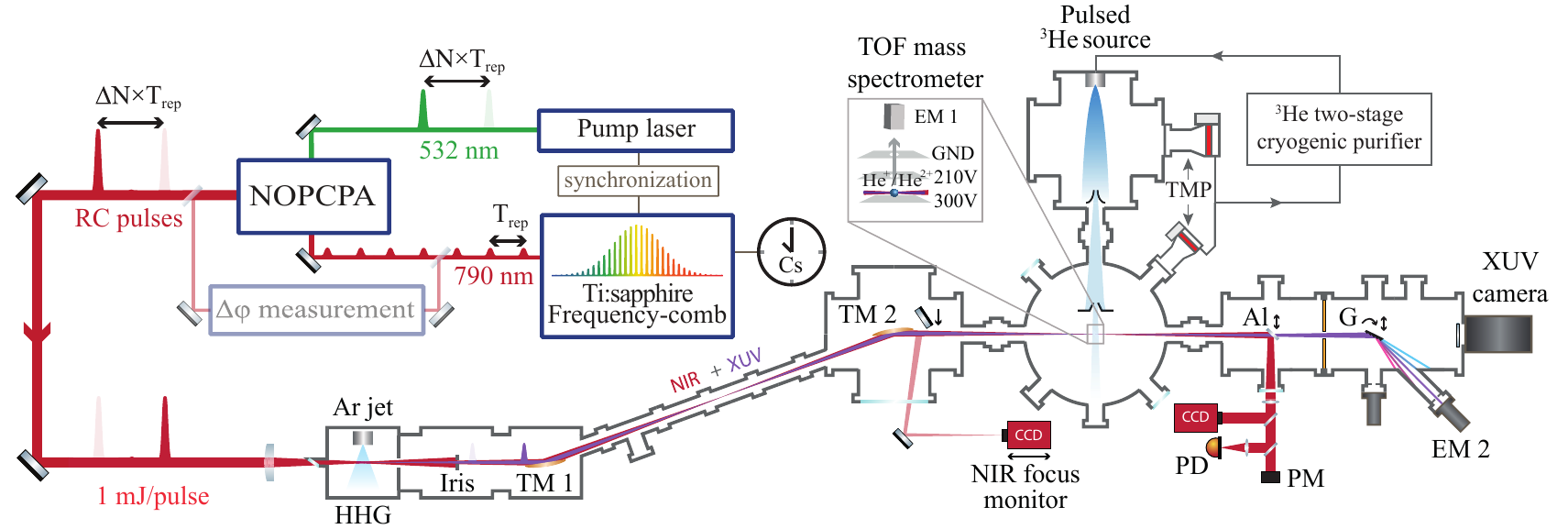}
\caption{\label{fig:vacuumsystem} Schematic of the laser and vacuum setup. A single pulse from a NIR Ti:sapphire frequency comb near 790\,nm is amplified to 1\,mJ in a NOPCPA, then upconverted to the XUV by HHG in an argon jet. The NIR pulse and the co-propagating harmonics are refocused together in a gas pulse of $^3$He using a pair of grazing-incidence toroidal mirrors (TM~1 and 2). A silver mirror after TM~2 can be moved into the beam to monitor the NIR focus outside vacuum. The ions formed after laser interaction are detected with a time-of-flight (TOF) mass spectrometer and electron multiplier (EM~1). A 400\,nm thick aluminium filter reflects the NIR beam, and transmits the XUV. The XUV is either monitored with an XUV camera or dispersed with a grating (G) to record the high harmonic spectrum with an electron multiplier (EM~2). PD stands for photodiode, PM for power meter and TMP for turbo-molecular pump. The system can produce two pulses, of which the relative phase can be measured after NOPCPA amplification for RCS (shown in gray, not used in this experiment).}
\end{figure*}
\indent The experimental setup is depicted in Fig.~\ref{fig:vacuumsystem}. The laser system consists of a Ti:sapphire FC laser, emitting a train of pulses (repetition rate $f_\text{rep}$ = 125\,MHz), from which two can be selectively amplified in a Noncollinear Optical Parametric Chirped Pulse Amplifier (NOPCPA) with a 28\,Hz repetition rate. For the experiment reported in this letter, we amplify only one pulse. The NOPCPA is specifically designed for high phase stability and is similar to the one used in Ref.~\cite{dreissen2020}, but then based on three lithium tri-borate (LBO) crystals that are pumped with a 532\,nm pump laser. The central wavelength of the amplified FC pulses can be varied between 785 and 800\,nm, and the spectral width is adjustable and set to 12\,nm (with a square spectral profile due to the NOPCPA). The output pulses of the NOPCPA have a duration of about 150\,fs (measured using the FROG technique~\cite{FROG}) after compression, close to the Fourier limit, and the pulse energy is approximately 1\,mJ.\\
\indent High harmonics of the amplified FC pulse are produced by focusing the beam to a spot of about 35\,\textmu m FWHM in a pulsed argon jet. The focus is placed slightly before the jet to achieve a minimally divergent harmonic beam by phase-matching only the short electron trajectories in the HHG process. The fundamental NIR pulses and the co-propagating harmonics are refocused together in a pulsed supersonic beam of helium, using a pair of grazing-incidence gold-coated toroidal mirrors. As the NIR beam diverges more strongly than the generated harmonics, an adjustable iris placed after the argon jet allows us to control the NIR intensity in the atomic beam independently of the harmonics. After the interaction region, a 400\,nm thick aluminum filter reflects the NIR light out of the vacuum system. This reflected light is used to monitor the pulse energy and stability. Wavelengths below $60$\,nm are transmitted by the aluminum filter and are either sent onto an Andor Newton CCD camera, or dispersed by an aluminium grating onto a Hamamatsu R5150-10 electron multiplier to measure the harmonic spectrum, see Fig.~\ref{fig:XUV}(a). With argon as the HHG medium, the cutoff wavelength is close to 32\,nm (the 25$^{\text{th}}$ harmonic). The 27$^{\text{th}}$ harmonic is strongly suppressed, and all higher harmonics are not observed.\\
\indent The He$^{+}$ and He$^{2+}$ ions are detected with a time-of-flight mass spectrometer, in which the arrival time on the ETP AF880 electron multiplier (EM) depends on the charge-to-mass ratio. By digitizing and integrating the EM trace with a DRS4 digitizer board, we record both the singly- and doubly-ionized helium signal for each laser shot. As $^4$He$^{2+}$ and H$_{2}^+$ ions have virtually the same charge-to-mass ratio, they cannot be distinguished in our detection system. This leads to a significant background signal from trace amounts of H$_2$ in the vacuum chamber and helium bottle. Therefore, we installed a cryogenic recycling system and performed all experiments with $^3$He. \\
\begin{figure}[b]
\includegraphics[width=0.48\textwidth]{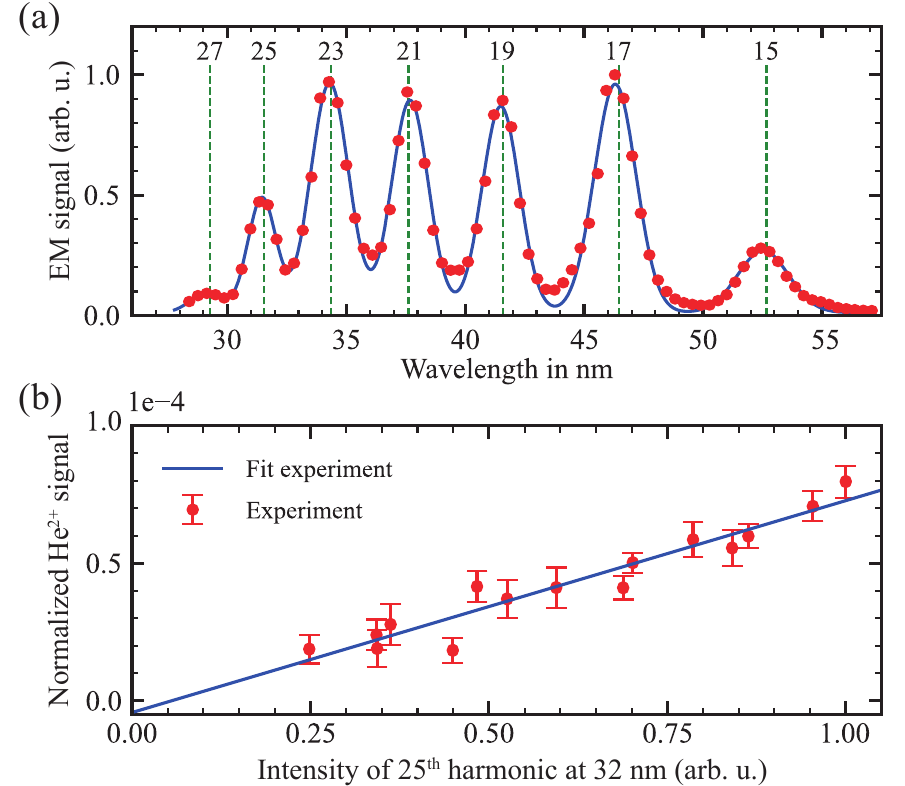}
\caption{\label{fig:XUV} (a) High harmonic spectrum recorded using the grating spectrometer (red dots) and fitted with multiple Gaussians (blue line).  The harmonic order is indicated for reference. Wavelengths above 50\,nm are suppressed due to an aluminum filter, and the width of each feature is instrument limited. (b) He$^{2+}$ signal as function of the 25$^{\text{th}}$ harmonic intensity, at a NIR central wavelength of 793\,nm and an intensity of 0.9$\times $10$^{14}$W/cm$^2$. The linear relation indicates that one photon at 32\,nm is involved in the excitation process.}
\end{figure}
\begin{figure*}
\includegraphics[width=0.98\textwidth]
{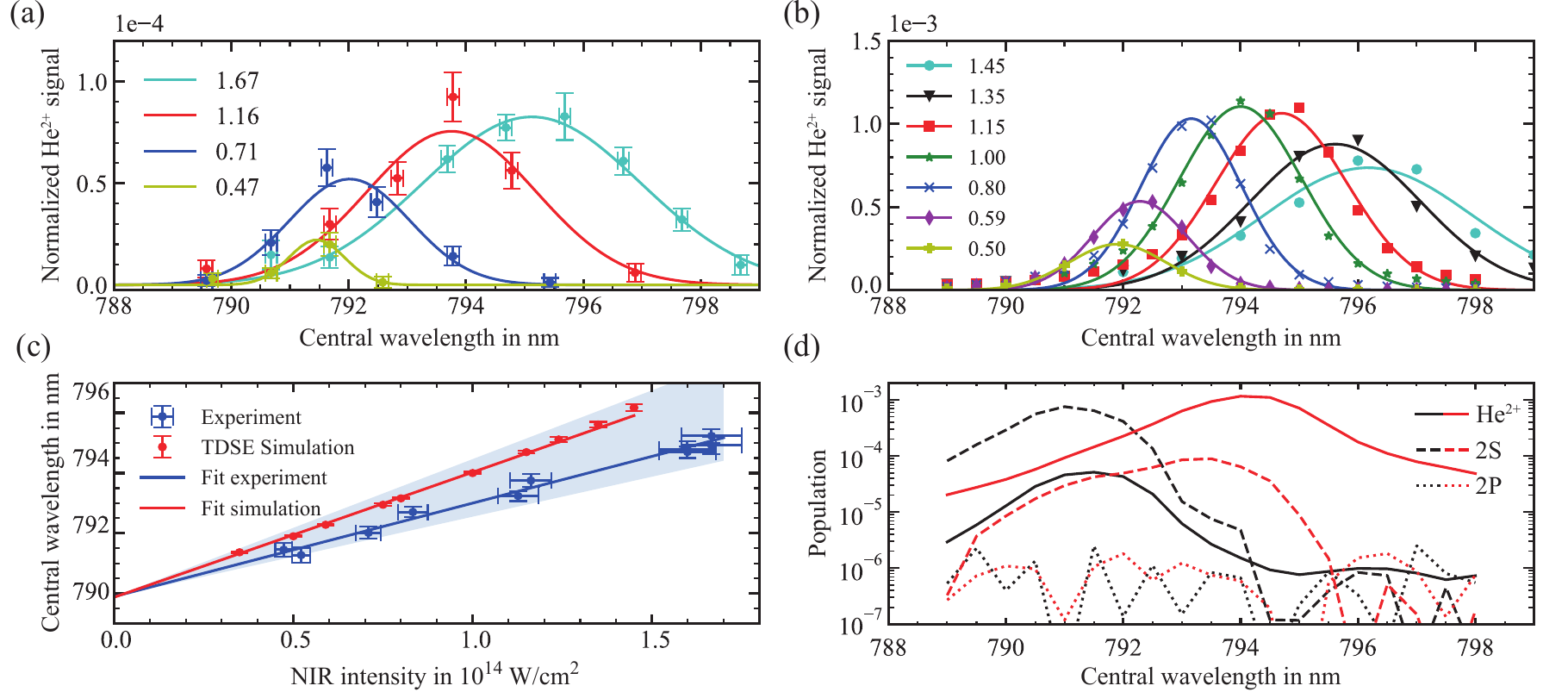}
\caption{\label{fig:AC_Stark_curves} (a) Normalized He$^{2+}$ signal as function of the NIR wavelength for several NIR intensities (in 10$^{14}$\,W/cm$^2$). This represents one of two recorded sets, see supplementary material. Gaussian fits are used to estimate the center, height, and width of the resonances. The measurement at the highest NIR intensity is an average of two scans. (b) TDSE simulation of the He$^{2+}$ yield, fitted with Gaussian functions, for several NIR intensities (in 10$^{14}$\,W/cm$^2$). (c) Fitted center of the resonances from the experiment and the TDSE simulations as a function of NIR intensity. The light-blue band represents the uncertainty of the estimation of the absolute NIR intensity, see Supplementary information. The extrapolated zero-intensity NIR resonant wavelength clearly agrees with the simulations and Ref.~\cite{NIST_ASD}, indicating excitation of the 1\textit{S}~--~2\textit{S} transition of He$^{+}$. (d) TDSE simulation of the normalized 2\textit{S}, 2\textit{P} and ion population at the end of the laser pulse for a NIR intensity of 0.35$\times$10$^{14}$\,W/cm$^2$ (black curves) and 1.0$\times$10$^{14}$\,W/cm$^2$ (red curves). On resonance, the 2\textit{S} state is populated three orders of magnitude more than the 2\textit{P}. At high NIR intensity, the 2\textit{S} population is reduced and the ionized fraction increased, identifying 1\textit{S}~--~2\textit{S} excitation as the main contribution to the ion signal.}
\end{figure*}
\indent To demonstrate excitation of the  1\textit{S}~--~2\textit{S} transition, we record the He$^{2+}$ signal (relative to the He$^{+}$ signal) as a function of the intensity and central wavelength of the NIR pulse, and as a function of the intensity of the 25$^{\text{th}}$ harmonic. A clear resonance is observed with a maximum excitation probability on the order of 10$^{-4}$ per laser pulse. The experimental results are consistent with our simulations based on numerically integrating the time-dependent Schr\"odinger equation (TDSE), and unambiguously exclude processes other than 1\textit{S}~--~2\textit{S} excitation, as explained below.\\
\indent In order to drive the 1\textit{S}~--~2\textit{S} transition and the nine-photon ionization step, NIR intensities higher than  3$\times$10$^{13}$\,W/cm$^2$ are required. At these intensities, the ac-Stark shift is a few tens of THz, pushing the resonance towards longer wavelengths. We use this feature to attribute the detected He$^{2+}$ signal to 1\textit{S}~--~2\textit{S} excitation and subsequent ionization. We record the He$^{2+}$ signal as a function of the NIR central wavelength for different intensities (see Fig.~\ref{fig:AC_Stark_curves}(a)), and use a linear extrapolation to extract the resonance position at zero intensity (Fig.~\ref{fig:AC_Stark_curves}(c)). The obtained zero-intensity NIR resonant wavelength of 789.95(18)\,nm is in close agreement with the value expected from the unperturbed spectrum of He$^{+}$ (789.877\,nm in combination with its 25$^{\text{th}}$ harmonic, see Fig.~\ref{fig:helium_levels})~\cite{NIST_ASD} and from our simulations that account for both the NIR and the (small) XUV ac-Stark shifts. This clearly points to 1\textit{S}~--~2\textit{S} excitation and excludes excitation of the nearby 1\textit{S}~--~2\textit{P} one-photon transition, of which the unperturbed excitation wavelength of 30.4\,nm corresponds to the 25$^{\text{th}}$ harmonic of 760\,nm. To further corroborate our interpretation of the data, we vary the 25$^{\text{th}}$ harmonic intensity by varying the argon density in the HHG gas jet, while keeping the NIR intensity constant. A clear linear relationship between the He$^{2+}$ signal and the 25$^{\text{th}}$ harmonic intensity is observed (see Fig.~\ref{fig:XUV}(b)), while the dependence on the other harmonics shows a nonlinear behavior (see Supplementary material). This is a strong indication that one photon of the 25$^{\text{th}}$ harmonic is involved in the excitation process, as expected from our two-photon excitation scheme shown in Fig.~\ref{fig:helium_levels}.\\
%
%
\indent As mentioned earlier, we compare the experimental data with simulations based on TDSE calculations~\cite{veniard1995,taieb1996,atomsInStrongFields}. Note that resonance enhancement based on two different driving wavelengths has been studied theoretically before in this manner, in the context of HHG in He$^+$~\cite{ishikawa2003}. In order to perform the TDSE calculations, we first calculate the harmonic field using the approach explained in Refs.~\cite{lewenstein_theory_1994,yakovlev2007,yudin2001}. Since this method only works well for harmonics in the plateau and cut-off region ($>$11$^\text{th}$), the lower harmonics are scaled to match the expected relative strengths~\cite{lopez2005}. The full harmonic spectrum is also scaled with a general factor to match an intensity of 2$\times$10$^9$\,W/cm${^2}$ for the 25$^\text{th}$ harmonic, which corresponds to the expected
experimental conditions in the interaction zone. The resulting XUV field, together with the fundamental NIR field, is used as input for our TDSE simulations. After integrating over the excitation pulses with a time step of 0.5~attoseconds, a projection of the electron wavefunction is made to identify the population of the bound states and to calculate the ionized (He$^{2+}$) fraction. The calculated yield of He$^{2+}$ is shown in Fig.~\ref{fig:AC_Stark_curves}(b). In the simulations, the maximum excitation probability is 10$^{-3}$ per pulse, which is an order of magnitude higher than seen in the experiment. This difference can be explained by the dynamics of the helium ionization process during the laser pulse, both spatial and temporal, which are not included in the simulations. Nevertheless, the simulations reproduce the features and trends observed in the experiment well within the experimental uncertainty (see also Supplementary Material).\\
\indent The TDSE calculations also enable us to follow the population of the relevant excited states, shown in Fig.~\ref{fig:AC_Stark_curves}(d) for two NIR intensities. The 2\textit{S} state is typically excited three orders of magnitude more than the 2\textit{P}. This is explained by the small spectral overlap of the HHG spectrum with the 1\textit{S}~--~2\textit{P} transition. Other states, such as the 3\textit{S} and 3\textit{P}, are excited even less than the 2\textit{P}. It can also be seen in Fig.~\ref{fig:AC_Stark_curves}(d) that at higher NIR intensity, a faster ionization to He$^{2+}$ leads to a rapid depletion of the 2\textit{S} population. Therefore, we attribute the spectral broadening and decrease of signal strength at the very highest NIR intensities to a reduction of the 2\textit{S} lifetime.\\
\indent The unequal wavelength scheme we demonstrated excludes the traditional method of first-order Doppler cancellation for a two-photon transition with counter-propagating laser beams. For a future RCS experiment, we propose an alternative by synchronizing the motion of a He$^+$ ion (the secular frequency) in a trap with the delay of the two excitation pulses in RCS. The ion will then be excited by both pulses at the same position in the trap, effectively cancelling the first-order Doppler effect.\\
\indent Another important aspect is the observed large ac-Stark shift of the 1\textit{S}~--~2\textit{S} resonance due to the NIR pulse. This might seem to limit precision spectroscopy, but with RCS the influence is much smaller because the resonance is only shifted during the two excitation pulses, and not in between. Moreover, a further reduction is achieved depending on how constant the pulse energy is kept as a function of the inter-pulse delay~\cite{RCStheoryanalysis}. With our RCS system, the (average) pulse energy is constant to 0.1\%, enabling spectroscopy with kHz-level accuracy (see Supplementary Information).\\
\indent In conclusion, we have observed the two-photon 1\textit{S}~--~2\textit{S} transition in $^3$He$^+$, based on excitation with 790\,nm and 32\,nm light derived from a frequency comb laser. A single-pulse excitation probability of 10$^{-4}$ was achieved, as well as a near-100\% ionization detection efficiency for the highest NIR intensities.
This demonstration represents an important step towards precision RCS of He$^+$ confined in an ion trap.\\
%
%

\indent We thank Rob Kortekaas and the members of the mechanical and electronic workshops for technical support. K.E. acknowledges support from the European Research Council via an ERC-Advanced grant (695677), and the Dutch Research Council (NWO) via grant 16MYSTP.\\

\bibliography{Grundeman_bibl}
\end{document}


\preprint{APS/123-QED}

\title{Supplementary Material}
\thanks{Belonging to: Laser Excitation of the 1\textit{S}~--~2\textit{S} Transition in Singly-Ionized Helium}

\author{E. L. Gr\"undeman}
\author{V. Barb\'e}%
\author{A. Mart\'inez de Velasco}
\author{C. Roth}
\author{M. Collombon}
\author{J. J. Krauth}
\author{L. S. Dreissen}
\affiliation{LaserLaB, Department of Physics and Astronomy, Vrije Universiteit, De Boelelaan 1081, 1081 HV Amsterdam, the Netherlands}
\author{R. Ta\"ieb}
\affiliation{Sorbonne Universit\'e, CNRS, Laboratoire de Chimie Physique-Mati\'ere et Rayonnement, LCPMR, F-75005 Paris Cedex 05, France
}
\author{K. S. E. Eikema}
\affiliation{LaserLaB, Department of Physics and Astronomy, Vrije Universiteit, De Boelelaan 1081, 1081 HV Amsterdam, the Netherlands}



\date{\today}

\maketitle

\section{Harmonic spectrum and XUV yield}
\begin{figure}
\includegraphics[width=0.48\textwidth]{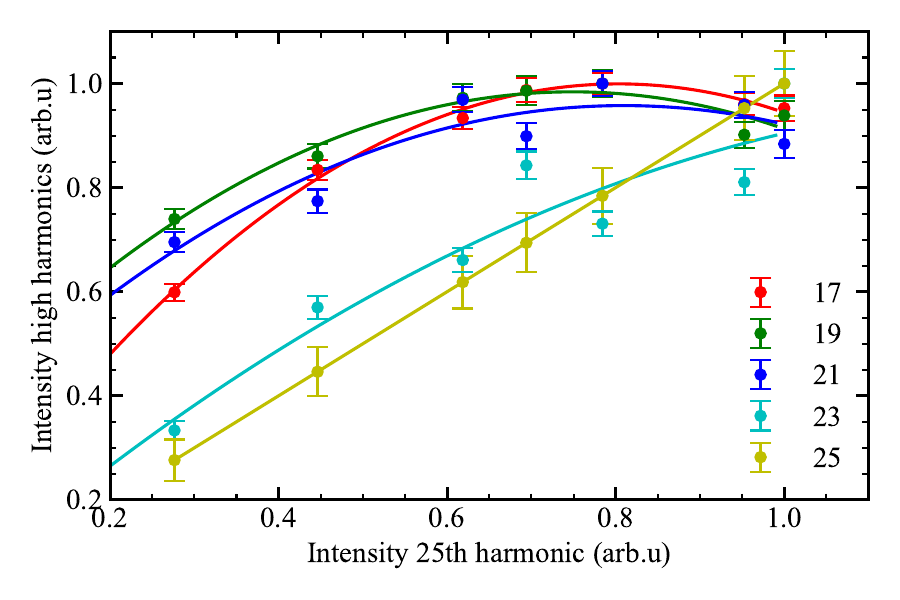}
\caption{\label{fig:Ratios_harmonics} Normalized intensities for different high harmonics as a function of the intensity of the 25$^\text{th}$ harmonic.}
\end{figure}

\begin{figure}
\includegraphics[width=0.48\textwidth]{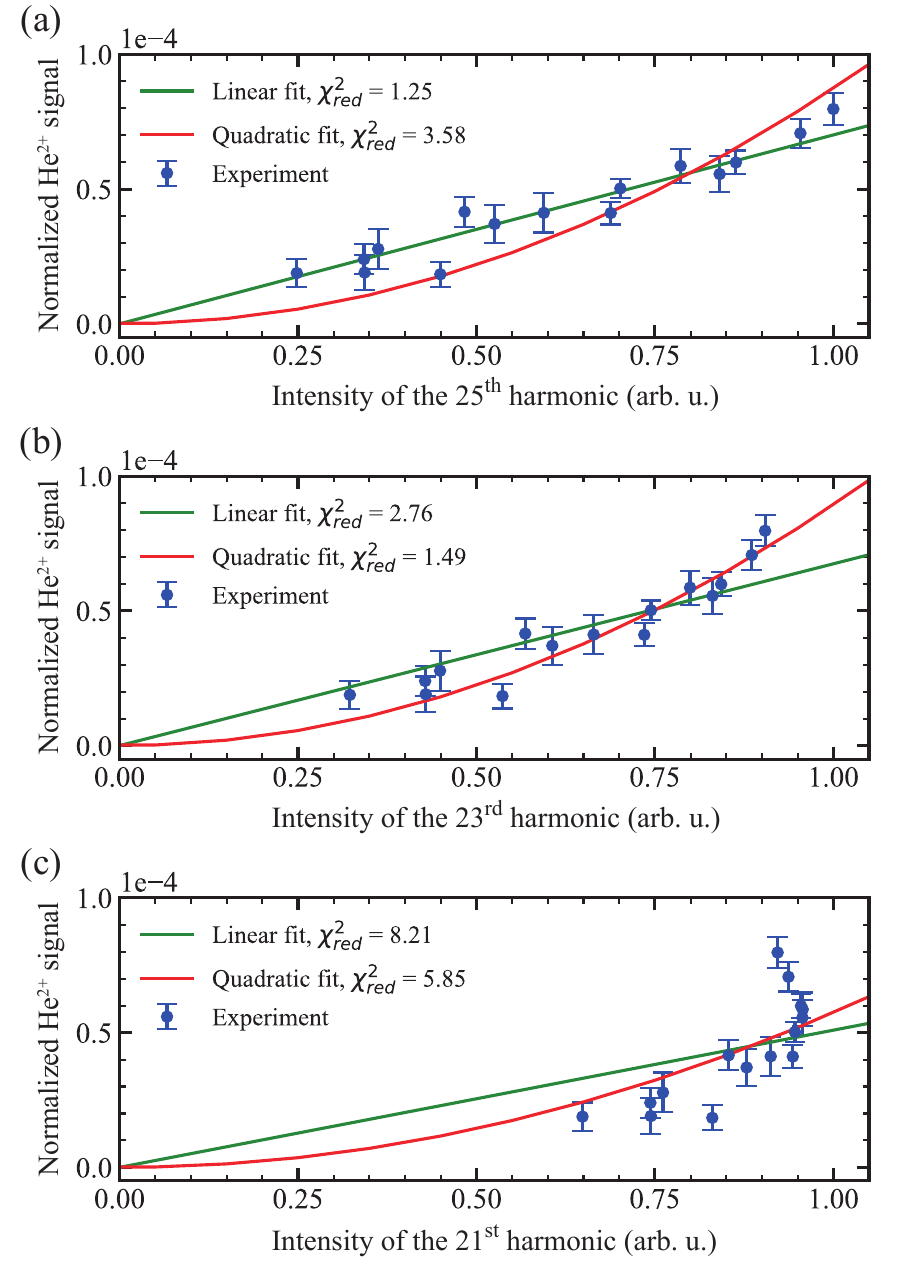}
\caption{\label{fig:harmonics_dependence} The normalized He$^{2+}$ signal dependence on the (a) 25$^{\text{th}}$, (b) 23$^{\text{rd}}$ and (c) 21$^{\text{st}}$ harmonic intensity. The reduced chi-squared for the linear and quadratic fits show the He$^{2+}$ signal is best described by a linear dependence on the 25$^{\text{th}}$ harmonic intensity.}
\end{figure}

\begin{figure}
\includegraphics[width=0.48\textwidth]{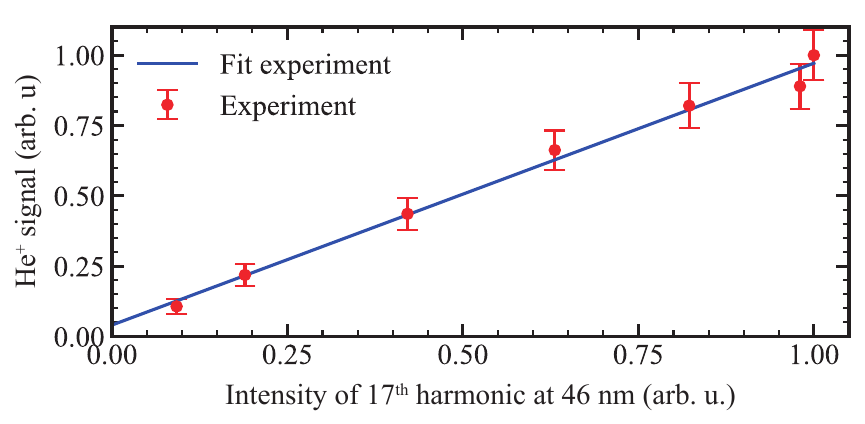}
\caption{\label{fig:He_ion_signal} The He$^+$ signal depends linearly on the intensity of the 17$^\text{th}$
harmonic at 46\,nm, indicating single-photon ionization.}
\end{figure}

\noindent We monitor the harmonic spectrum with a grating spectrometer, as shown in Fig.~3(a) in the main text. The spectrometer consists of a grating (model
33009FL01-510H from Newport) and an electron multiplier (Hamamatsu R5150-10). The grating efficiency is not specified for our wavelength range (30 - 60\,nm), therefore we cannot use our measurement for an absolute determination of our XUV pulse energies. Instead, we use the spectrum to monitor the stability of individual harmonics during experiments and to keep the conditions stable between experiments.\\
\indent In the light field many harmonics are present, which could all interact with the He$^+$ ion and lead to the observed ionization. By varying the HHG conditions we can disentangle the influence of the different harmonics on the signal. Since the highest harmonic generated (the 25$^{\text{th}}$) is close to the cutoff wavelength, it is much more sensitive to the argon density in the HHG gas jet. Therefore, we can vary the Ar backing pressure and correlate the He$^{2+}$ signal with the intensities of the different harmonics. As shown in Fig.~3(b) in the main text, the observed He$^{2+}$ signal is linear with the intensity of the 25$^{\text{th}}$ harmonic, as is expected from our excitation scheme. During this experiment we monitored only the 25$^{\text{th}}$ harmonic intensity, and changed the backing pressure (in a random fashion) to investigate the influence on the signal. After recording the ion signal, we investigated the harmonic spectra for different argon pressures. From a multi-Gaussian fit we find the relative strengths of all the harmonics for the different argon pressures. In Fig.~\ref{fig:Ratios_harmonics}, the normalized intensities of the other harmonics as a function of the 25$^{\text{th}}$ harmonic are shown, with a polynomial fit to estimate the relative ratios.\\
\indent We use the fits in Fig.~\ref{fig:Ratios_harmonics} to investigate the dependence of the He$^{2+}$ signal on the lower harmonics, as shown in Fig.~\ref{fig:harmonics_dependence}. Given the relatively low HHG photon flux, only processes involving one or two photons are taken into account as a possible route to ionize He$^+$ from the ground state, so only linear and quadratic fits are shown. From the reduced $\chi^2$ it follows that linear dependence on the 25$^{\text{th}}$ harmonic intensity gives the best fit, followed by a quadratic fit of the 23$^{\text{rd}}$ harmonic intensity. However, a two-photon process involving  the 23$^{\text{rd}}$ harmonic would be non-resonant, as there are no states nearby. This route is ruled out by the NIR central wavelength scans, which show a clear resonance. Therefore, the most probable route to He$^{2+}$ signal is a process involving one photon of the 25$^{\text{th}}$ harmonic.\\
\indent We use a similar approach to determine the ionization mechanism of neutral helium. The 17$^\text{th}$ harmonic at 46\,nm is the lowest harmonic that is above the ionization threshold of 24.6\,eV, and is expected to be the main contribution to the ionization. The He$^+$ signal on the electron multiplier depends linearly on the intensity of the 17$^\text{th}$ harmonic (see Fig.~\ref{fig:He_ion_signal}), indicating single photon ionization. For the highest 17$^\text{th}$ harmonic intensity used in this experiment, we detect an estimated 150 He$^+$ ions per shot.

\section{Comparison experiment and theory}
\begin{figure}
\includegraphics[width=0.48\textwidth]{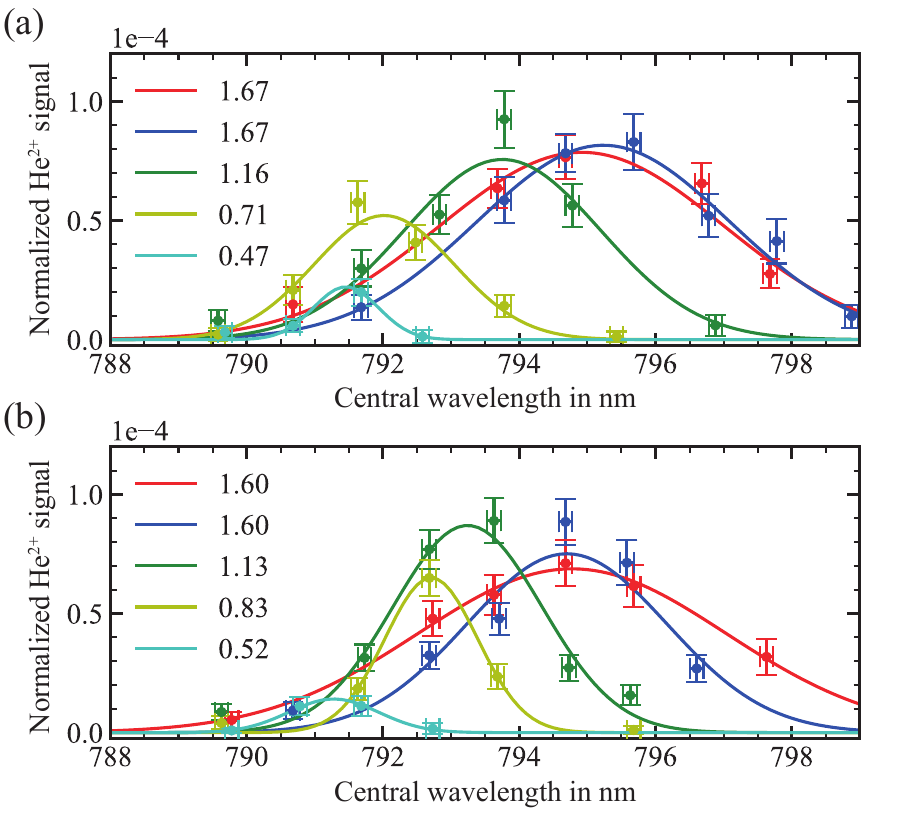}
\caption{\label{fig:Experimental_data} Normalized He$^{2+}$ signal as function of the NIR wavelength for several NIR intensities (in 10$^{14}$\,W/cm$^2$). Gaussian fits are used to estimate the center, height, and width of the resonances. Set (a) and (b) are two different measurement days. Set (a) is shown in the main text, with the two curves at the (same) highest intensity averaged together.}
\end{figure}

\begin{figure}[b]
\includegraphics[width=0.48\textwidth]{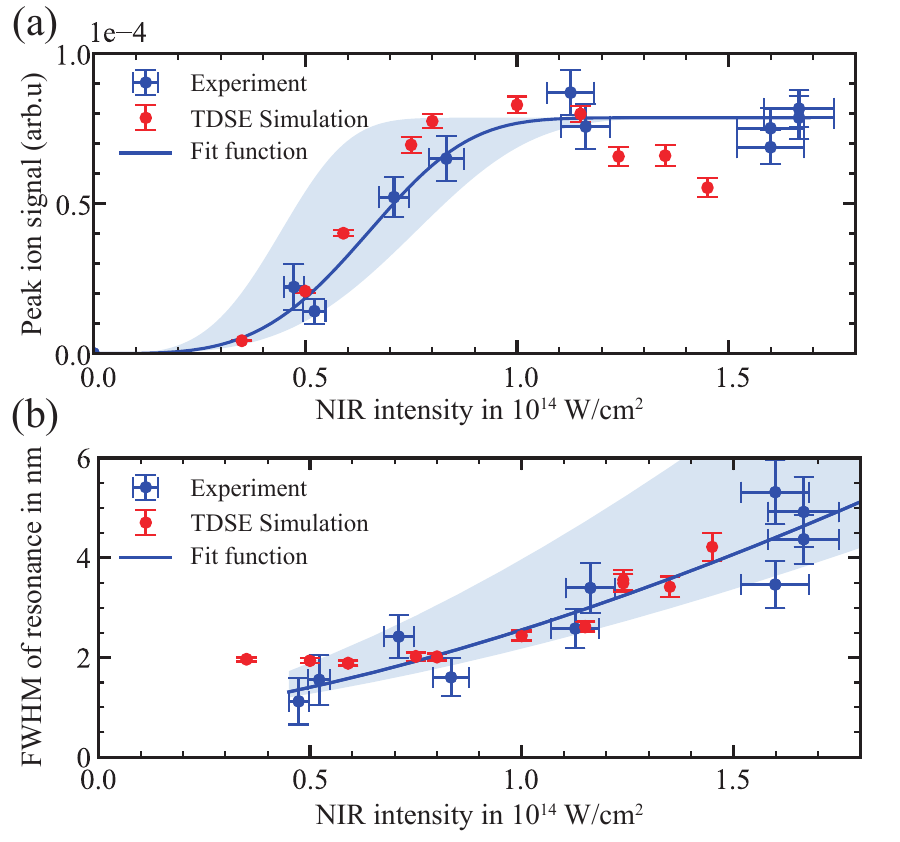} 
\caption{\label{fig:height_width} (a) Extracted amplitude (height) of the resonances of the experimental data compared to the TDSE simulations as a function of NIR intensity. The theoretical values have been scaled with a general factor to match the experimental values at high intensity. A generic multiphoton ionization fit function $a(1-\textrm{e}^{-b\,x^c})$ is fitted to the experimental data and displayed in blue. (b) Width of the resonances as a function of NIR intensity. A generic fit function $a+b\,x^c$ is fitted to the experimental data and displayed in blue. For both (a) and (b) the vertical error bars associated with the experimental data are statistical error bars. A confidence interval corresponding to the experimental systematic uncertainties (see text) is also displayed in light blue along the corresponding fit function.}
\end{figure}

Two complete measurement sets were recorded, on two separate days, by scanning the NIR wavelength and the power of the NIR pulses, as shown in Fig.~\ref{fig:Experimental_data}. All the extracted center wavelengths are included in Fig.~4(c) in the main text for the extrapolation to zero intensity. A further comparison with the simulations can be made by comparing the width and heights of the observed resonances, as shown in Fig.~\ref{fig:height_width}. The general shapes of the measured and calculated peak heights are very similar, where the calculated peak heights are scaled by a general factor to match the experiment (see Fig.~\ref{fig:height_width}(a)). As for the widths, the simulations give a width that is almost constant until the intensity becomes higher than 1.0$\times$10$^{14}$\,W/cm$^2$, after which it increases with intensity. We observe a similar trend experimentally with a clear increase of the resonance width at the highest intensities. At low intensities, the experimental data seem to indicate an increasing linear trend rather than a constant width. Because we could only sample a few wavelengths for the NIR in the available measurement time, it led to fewer points on the low power resonance features than for the broadened resonance at high power. Therefore, the indicated vertical error bar might be an underestimation of the measurement error. A deviation of a single point for the narrower curves can result in a significant width change, so that it is difficult to draw definite conclusions in this case.\\
\indent We estimate the NIR intensities using the measured NIR power, pulse length and shape, and focus size. The pulse length is measured using Frequency Resolved Optical Gating (FROG) before the vacuum system and is close to the Fourier limit for a square spectrum. The focus size is measured by raising a silver mirror into the beam after the second toroidal mirror, reflecting the NIR beam out onto a camera. We use multi-photon ionization of trace amounts H$_2$O in our helium gas pulse as a sensitive tool to optimize the NIR focus position in the atomic beam. The NIR power is measured with a power meter at the end of the vacuum system. We assume the interaction happens only when the 25$^\text{th}$ harmonic is present, which is during the central 30\,fs of the 150\,fs NIR pulse. Additionally, it is assumed that the XUV focus is much smaller than the NIR focus and that they are overlapped, such that the excitation takes place only in the most intense part of the NIR beam. The intensities shown in the figures are calculated under these assumptions. If some of these assumptions are not correct, the intensity experienced by the He$^+$ ions will be different by a constant factor. Most of the possible errors, e.g. the pulse not being Fourier-limited or being outside the focus, will only lead to an overestimation of the NIR intensity in the interaction region. We estimate the possible errors to be as follows: an up to $10\%$ longer NIR pulse, up to $5\%$ larger focus size, a $\pm\,10\%$ error on the power meter reading and an up to $5\%$ higher than expected optical loss in the vacuum outcoupling of the NIR. These systematic uncertainties are accounted for by the confidence intervals displayed in light blue in Fig.~4(c) in the main text and in ~Fig.~\ref{fig:height_width} in the Supplementary material.
\section{Estimation transition probability}
In the simulations the helium ion is prepared in the 1S state, and then interacts with the laser field as if it were in the focus. This is a good representation of the future situation where the ion is confined in the Paul trap. However, for the beam experiment the ions are dynamically created during the same pulse in which the excitation takes place. These ions are created by the 17$^\text{th}$ harmonic, which has a larger focus spot size and a longer pulse length than the 25$^\text{th}$.\\
\indent The ionization rate of neutral helium is linear with the 17$^\text{th}$ harmonic intensity, therefore most ions are produced at the peak of the pulse. These ions can only interact with the part of the 25$^\text{th}$ harmonic pulse that is still to come. A general formula for the average 25$^\text{th}$ harmonic pulse energy that the ions created by the 17$^\text{th}$ harmonic will interact with is given by:
\begin{align}
\frac{\text{Log}16}{\pi\tau_{17}\tau_{25}} \int_{-\infty}^\infty \left(e^{-4\text{Log}2 \frac{t^2}{\tau_{17}^2}}\int_t^\infty e^{-4\text{Log}2\frac{t^2}{\tau_{25}^2}}\right)\text{d}t = \frac{1}{2}
\end{align}
This outcome is independent of the pulse lengths (as long as the timing of the pulses coincides). The created ions can also only interact with on average half the NIR intensity, which means the excitation probability decreases by another 50\%.\\
\indent The second assumption in the simulation is that the interaction with He$^+$ happens in the focus of the laser. In reality the ions are created in a region which depends on the focus size of the 17$^\text{th}$ harmonic, and the interaction happens in a region defined by the focus size of the 25$^\text{th}$harmonic. The average intensity of the 25$^\text{th}$ harmonic that the ions experience is given by:
\begin{align}
\frac{\text{Log} 16}{\pi \omega_{17}^2}\int_{-\infty}^\infty\int_{-\infty}^\infty e^{-4 \text{Log}2\frac{x^2 + y^2}{\omega_{17}^2}} e^{-4 \text{Log}2\frac{x^2 + y^2}{\omega_{25}^2}} \text{d}x\,\text{d}y
\end{align}
For the estimated spot sizes of $\omega_{17} = 10$\,\textmu m and $\omega_{25} = 8$\,\textmu m, this average intensity is 40\% of the peak intensity. Combined with the temporal effect described before, we expect only 10\% of the simulated He$^{2+}$ yield in the beam experiment. This is close to the scaling factor of 0.075 applied to the simulated heights in Fig.~\ref{fig:height_width}a to make comparison between the theory and experiment.\\
\noindent This now allows us to make an estimate of the expected Ramsey-comb signal. With the current laser configuration the expected single pulse excitation probability on a trapped ion will be close to 10$^{-3}$. For two pulses as in RCS, the peak of the signal fringe will be a factor 4 higher due to constructive interference of the two excitation contributions. Also the HHG yield can be higher than in the current experiment (by at least a factor 2, as we found out after the measurements were completed), leading to an expected peak excitation probability of 10$^{-2}$. With an experimental repetition rate of 28\,Hz, this would yield one excitation every 3.5\,s on the peak of the fringe.
\section{ac-Stark shift estimation on RCS}
We observe a large ac-Stark shift of the transition at the NIR intensities that we use (see Fig.~\ref{fig:Experimental_data}). In RCS this ac-Stark shift does not influence the extracted transition frequency, provided the relative energy of the pulses is kept constant for all interpulse delays. If, on the other hand, the relative (average) energy of the two RCS pulses is slightly different between the recorded Ramsey fringes at different inter-pulse delays, this leads to an error in the extracted transition frequency. We estimate below the magnitude of the associated error expected in our future RCS experiment.\\
\indent If we use double ionization for detecting the He$^+$ 2S state, as we do in our present experiment, we could employ a separate ionization pulse after the two-pulse RCS sequence. Then we can choose a NIR power that optimizes 1S-2S excitation, but keeps the ionization per RCS pulse below 10\%. 
This corresponds to the lowest NIR intensity of 0.35$\times$10$^{14}$\,W/cm$^2$ we used in the current experiment. In this case the ac-Stark shift is about 20\,THz (see Fig.~4(a) of the main text). For RCS, it is the accumulated phase shift of the excited superposition that matters. We can estimate this if we take the excitation window of our current experiment equal to the pulse length of the 25$^\text{th}$ harmonic (which we estimate at 27\,fs pulse width from our HHG simulation), and calculate the effect of the NIR exposure during a full RCS sequence. 
The observed 20\,THz ac-Stark shift in the current experiment during 27\,fs is equivalent to an average phase shift of 3.3\,rad. Please note that the phase shift depends on the intensity, but what matters is the integrated effect over the pulse, which also means that it scales with the integrated intensity, and so the pulse energy.\\
In the current experiment, the atoms are exposed to only 17\% of the total energy of the NIR pulse during two-photon excitation. In RCS, the equivalent of one full NIR pulse is causing a phase shift (roughly equal to the sum of the second half of the first RCS pulse, and the first half of the second RCS pulse). 
This amounts to 20\,rad at 0.35$\times$10$^{14}$\,W/cm$^2$. Based on this value there are two effects to consider. Fluctuations in pulse energy can lead to phase noise, and therefore contrast loss of the recorded Ramsey signal, but not to a systematic shift. Given a NIR pulse stability of approximately 1\%, we expect a phase noise at the transition frequency of 200 mrad, which is small enough to have almost no effect. As mentioned earlier, the other aspect is the stability of the pulse energy for different pulse delays of our RCS laser system, as this could give a systematic phase shift difference. We commonly achieve a pulse energy stability as a function of delay of 0.1\%. Therefore this variation could lead to an effective systematic phase shift of about $\Delta\Phi = 20$\,mrad. The resulting error in the frequency determination with RCS is then $\Delta\Phi/(2\pi\, T)$, where $T$ is the maximum difference in inter-pulse delay used in the RCS sequence. The delay that can be employed is influenced by factors such as sources of decoherence and the maximum pulse delay of 10 microsecond of the laser system; the natural lifetime of the 2\textit{S} state of He$^{+}$ is 1.9\,ms and will not be limiting. For the first experiments we aim at $\Delta T$~$>$~400\,ns, resulting in an uncertainty (error) of less than 10\,kHz from the ac-Stark effect. To reach sub-kHz uncertainty on the ac-Stark shift we will then aim for $T$~$>$~3\,$\mu$s in the next stages of the experiment. These numbers indicate that, despite the large instantaneous ac-Stark shift during the laser pulses, kHz-level RCS on the 1\textit{S}~--~2\textit{S} two-photon transition in singly-ionized helium is feasible with the demonstrated excitation method. 